# Mozart Effect, Cognitive Dissonance, and the Pleasure of Music

Leonid Perlovsky[1,2], Arnaud.Cabanac[3], Marie-Claude Bonniot-Cabanac[4], Michel Cabanac[4]

[1]Harvard University, [2]AFRL, [3]De Rochebelle School (C.S.D.D), Quebec, [4]Faculty of Medicine, Laval University, Quebec, Canada

Abstract

The 'Mozart effect' refers to scientific data on short-term improvement on certain mental tasks after listening to Mozart, and also to its popularized version that "listening to Mozart makes you smarter" (Tomatis, 1991; Wikipedia, 2012). Does 'Mozart effect' point to a fundamental cognitive function of music? Would such an effect be due to the hedonicity, a fundamental dimension of mental experience? The present paper explores a recent hypothesis that music helps to tolerate cognitive dissonances and thus enabled accumulation of knowledge and human cultural evolution (Perlovsky, 2010, 2012). We studied whether the influence of music is related to its hedonicity and whether pleasant or unpleasant music would influence scholarly test performance and cognitive dissonance. Specific hypotheses evaluated in this study are that during a test students experience contradictory cognitions that cause cognitive dissonances. If some music helps to tolerate cognitive dissonances, then first, this music should increase the duration during which participants can tolerate stressful conditions while evaluating test choices. Second, this should result in improved performance. These hypotheses are tentatively confirmed in the reported experiments as the agreeable music was correlated with better performance above that under indifferent or unpleasant music. It follows that music likely performs a fundamental cognitive function explaining the origin and evolution of musical ability that have been considered a mystery.

## 1. Introduction - Mozart effect, cognitive dissonance, and music

The 'Mozart effect' is a short-term improvement on "spatial-temporal reasoning" (Tomatis, 1991). The idea that "listening to Mozart makes you smarter" (Wikipedia, 2012) has been so much hyped by the media (> 500,000 sites on Google) that many scientists conducted experiments to verify its validity. A short-term effect of any improvement was illustrated, and specificity to Mozart and music was questioned (Steele et al 1999; Thompson et al 2001; Schellenberg, 2006). Here we explore the 'Mozart effect' as a probe into the possible fundamental cognitive function of music.

Music, its strong power over humans, its origin and cognitive function have been a mystery for long time. Aristotle (1995) listed the power of music among the unsolved problems. Kant (1790), explaining the epistemology of the beautiful and the sublime, could not explain music: "(As for) the expansion of the faculties... in the judgment for cognition, music will have the lowest place among (the beautiful arts)... because it merely plays with senses." According to Darwin (1881), the human musical faculty "must be ranked amongst the most mysterious with which (man) is endowed" because music is a human cultural universal that appears to serve no obvious adaptive purpose. Among current evolutionary psychologists some argue that music plays no adaptive role in human evolution. So following Kant, Pinker (1997) has suggested that music is an "auditory cheesecake," a byproduct of natural selection that just happened to "tickle the sensitive spots." Other contemporary scientists suggest that music clearly has an evolutionary role, and point to music's universality (Masataka, 2009). In 2008, Nature published a series of essays on music (Editorial, 2008). The authors agreed that music is a cross-cultural universal, still "none... has yet been able to answer the fundamental question: why does music have such power over us?" (Ball, 2008).

Cognitive dissonance (CD) is "a discomfort caused by holding conflicting cognitions" (Festinger 1957; Wikipedia 2012). It is known that this discomfort is usually resolved by devaluing-discarding a conflicting cognition. Any two elements of knowledge contradict each other to some extent, leading to CD and to de-motivation of knowledge accumulation. In particular, emergence of language and the following accumulation of knowledge would be devalued. Therefore the current theory of CD questions motivations for the entire human evolution, unless a powerful cognitive mechanism would emerge in parallel with language, which would enable keeping in mind contradictory cognitions.

A recent hypothesis suggests that music originated in human evolution to help overcoming negative consequences of CD (Perlovsky, 2010, 2012). Music was argued to be this powerful mechanism overcoming negative effects of CD. Because all decisions are made in the hedonic dimension of consciousness (Bonniot-Cabanac & Cabanac, 2009) and result from the maximization of pleasure (Cabanac & Bonniot-Cabanac 2007), we may suspect that such an influence of music also takes place in that dimension, *i.e.* that pleasure/displeasure operates also in the case of the Mozart effect.

As pleasure (Cabanac 1992) was shown to be the 'common currency' postulated by McFarland and Sibly (1975) to allow motivations to 'talk' to one another and establish behavioral priorities, it was natural to explore whether pleasure would also fulfill the same function with CD where two cognitions are in conflict with one another in our mind. A music ability to help keep contradictory cognitions in mind has been demonstrated experimentally in (Masataka & Perlovsky 2012): music has helped young children (4 y.o.a.) to avoid devaluing an attractive toy, while not playing with it. The fundamental and broad claims about musical role in cognition and human evolution require multifaceted evaluation. Here we approach relations between music and CD in a different setting of student performance on academic tests. We evaluate two hypotheses: first, that the hedonicity (*i.e* pleasure or displeasure) from music could modulate the ability to tolerate stress caused by CD, and second, whether the result would lead to an applied use of music during academic examination tests.

## 2. Methods

Two groups of 5th year high school (14-15 y.o.a.) of both sexes served as participants. They answered a multiple choice type training test with 12 questions of their scientific course for fifteen minutes. After they had completed the test each received a form with 3 additional questions on face A and, after answering those, two final questions on face B (reverse).

Environmental music
What was aimed at was to play music to both groups: one calm and quiet to one group and, to the second group, a music widely different, vivid, and drawing attention. The nature and hedonicity of the environmental music played during the test had been selected by probing on other teenagers who did not participate in the experiment. Calm music was Mozart sonata in D for two pianos K.448, (used by Masataka et al., 2012) especially the Andante; it was determined to be 'Pleasant', and the other music was a koto solo with some disharmonious sequences, by Kuro Kami and Sakura Miyotote, determined to be 'UnPleasant'. The loudspeaker had been placed near the ceiling in the center of the class-room. In both sessions the music intensities were: 55 dB homogenous in the room, as checked from a sonometer.

Participants

Sixty four participants took place in two groups and subjected to Mozart (n=32) and Koto (n=32) music. The two groups had identical grades performance.

It happened that some of the participants in the Mozart group, found that music UnPleasant, and some from the koto group found it Pleasant. Therefore the results were sorted, not on the account of the music heard, but on the pleasure or the displeasure experienced: 30 rated their music as Pleasant, and 21 as UnPleasant. Also, in both groups (13 participants altogether), rated on Questionnaire page B (see below) the music they heard as indifferent (zero hedonicity). These 13 participants served as a control and were labeled the 'UnHedonic' group.

Questionnaires

At the end of their academic tests the participants answered two short questionnaires on separate pages, the first one, Page A, probed their behavioral performance and the second one, Page B, probed their experience. This protocol was arranged that way in order to avoid drawing the participants' attention on the environmental music and on their awareness aroused by the previous questions on face A. The music had been stopped at the end of the test, i.e. before these final questionnaires were opened.

On Page A the participants were requested to:
- write the exact time of their completion of the academic test; thus providing their individual duration
- rate from 0 to 10 how difficult they had found the test
- write from 0 to 100 the grade they expected to have earned
- rate from 0 to 10, the intensity of their stress

On Page B the participants answered the following two questions:
- Have you been aware that music was played during the test? Answer Y/N
- Did you like it? Rate your pleasure/displeasure experience, as a number between -5 and +5, with the following landmarks: -5 very unpleasant, -3 unpleasant, 0 indifferent, +3 agreeable, +5 very agreeable.

## 3. Results

### 3.1 Grade performance and hedonicity of music

Grades earned under Pleasant music condition were higher than for UnPleasant or UnHedonic. The differences are statistically significant. (This is similar to the 'Mozart effect'); Relationships for UnPleasant vs. UnHedonic are not statistically significant, Table 1. Here and below MW and T denote Mann-Whitney and Student T-tests; arrows <-...-> indicate the compared pair of conditions; reported numbers for each pair of conditions show the probability of accepting null hypothesis, p (p = 1 corresponds to no difference, $p < 0.05$ is usually interpreted as statistically significant difference between the pair of conditions).

**Table 1.** Grades earned (from 0 to 16). Significant improvement for Pleasant music condition above the two other groups. Higher grades under pleasant music than under unpleasant or indifferent music, were statistically significant.

|                          | Pleasant                          | UnPleasant                          | UnHedonic          |
| ------------------------ | --------------------------------- | ----------------------------------- | ------------------ |
| median                   | 14.00                             | 12.00                               | 12.50              |
| mean                     | 13.60                             | 12.17                               | 12.12              |
| st.dev.                  | 1.84                              | 2.39                                | 2.58               |
| Pleasant > UnPleasant    | <-- $p_{MW} = 0.02$; $p_T = 0.01$-->       |                                     |                    |
| Pleasant > UnHedonic     | <--------------- $p_{MW} = 0.03$; $p_T = 0.04$--------------->  |                    |
| UnPleasant vs. UnHedonic |                                   | <-- $p_{MW} = 0.96$; $p_T = 0.48$ -->        |                    |

## 3.2 Other variables

**Table 2.** Duration was shorter under Pleasant music than under UnHedonic music and UnPleasant condition. The difference did not reach statistical significance for Pleasant vs. UnPleasant, but significant for Pleasant vs. UnHedonic.

|                          | Pleasant                          | UnPleasant                          | UnHedonic          |
| ------------------------ | --------------------------------- | ----------------------------------- | ------------------ |
| median                   | 11.00                             | 11.00                               | 13.00              |
| mean                     | 11.13                             | 11.76                               | 12.69              |
| st.dev.                  | 2.43                              | 2.23                                | 2.10               |
| Pleasant > UnPleasant    | <-- $p_{MW} = 0.44$; $p_T = 0.17$-->       |                                     |                    |
| Pleasant > UnHedonic     | <--------------- $p_{MW} = 0.05$; $p_T = 0.02$--------------->  |                    |
| UnPleasant vs UnHedonic  |                                   | <-- $p_{MW} = 0.22$; $p_T = 0.12$ -->        |                    |

**Table 3.** Rating for Difficulty. Ratings were lower for Pleasant condition than for the other two. These differences do not reach the threshold for statistical significance ($p < 0.05$).

|                          | Pleasant                          | UnPleasant                          | UnHedonic          |
| ------------------------ | --------------------------------- | ----------------------------------- | ------------------ |
| median                   | 4.00                              | 5.00                                | 5.00               |
| mean                     | 4.35                              | 4.50                                | 4.92               |
| st.dev.                  | 2.00                              | 2.22                                | 2.43               |
| Pleasant > UnPleasant    | <-- $p_{MW} = 0.80$; $p_T = 0.40$-->       |                                     |                    |
| Pleasant > UnHedonic     | <--------------- $p_{MW} = 0.43$; $p_T = 0.23$--------------->  |                    |
| UnPleasant vs UnHedonic  |                                   | <-- $p_{MW} = 0.60$; $p_T = 0.28$ -->        |                    |

**Table 4.** Median Expected_Grade was higher for Pleasant condition than for the other two. These differences are statistically significant.

|                          | Pleasant                          | UnPleasant                          | UnHedonic          |
| ------------------------ | --------------------------------- | ----------------------------------- | ------------------ |
| median                   | 80.00                             | 75.00                               | 75.00              |
| mean                     | 80.70                             | 74.62                               | 70.85              |
| st.dev.                  | 9.90                              | 9.67                                | 15.49              |
| Pleasant > UnPleasant    | <-- $p_{MW} = 0.04$; $p_T = 0.02$-->       |                                     |                    |
| Pleasant > UnHedonic     | <--------------- $p_{MW} = 0.04$; $p_T = 0.03$--------------->  |                    |
| UnPleasant vs UnHedonic  |                                   | <-- $p_{MW} = 0.73$; $p_T = 0.22$ -->        |                    |

**Table 5.** Stress rating was lower for Pleasant condition than for the other two. These differences are of low statistical significant for Pleasant vs. UnPleasant conditions, and of much lower (then p=0.05) statistical significance for the other two pairs of conditions.

|  | Pleasant | UnPleasant | UnHedonic |
|---|---|---|---|
| median | 2.50 | 4.00 | 4.00 |
| mean | 3.07 | 4.48 | 4.00 |
| st.dev. | 2.36 | 2.66 | 2.68 |
| Pleasant > UnPleasant | <-- $p_{MW} = 0.08$; $p_T = 0.03$--> | | |
| Pleasant > UnHedonic | <--------------- $p_{MW} = 0.35$; $p_T = 0.14$---------------> | | |
| UnPleasant vs UnHedonic | | <-- $p_{MW} = 0.79$; $p_T = 0.31$ --> | |

We would add that many pair wise correlations are significant, and their signs are, as expected, consistent with our hypotheses, without adding new unexpected information. For example, correlation of difficulty and duration over conditions is positive (0.163) as expected: more difficulty implies more time needed to answer test questions. In the next section we will address why this value is of relatively low significance.

3.3 Evaluation of the hypothesis that the hedonicity of music modulates the tolerance for cognitive dissonance (**Table 6** below)

To isolate the effect of CD-stress on reducing durations, we compute regression of Duration on two variables, the 1st measuring difficulty for each student (estimated as either subjective Difficulty, or Expected grade, or Grade) and the 2nd Stress. To get data independent of units of measurements, we consider normalized variables, mean values are subtracted for every variable and the results are divided by standard deviations. For normalized variables regression equation looks like follows (Anderson, 1984):

Duration = a1*"Difficulty" + a2*Stress

The advantage of using normalized variables is that results are independent of units of measurements of individual variables. Coefficient a1 gives a dimensionless isolated effect of "Difficulty" on Duration, and coefficient a2 gives a dimensionless isolated effect of Stress on Duration. "Difficulty" can be estimated as (subjective Difficulty), or as (-Grade), or as (-Expected Grade). A higher grade measures "easiness" rather than "difficulty", therefore to measure "difficulty" we took negative values (-Grade, or –Expected Grade). We computed all three regressions for each condition, and evaluated statistical significance of the effects of "difficulty" and stress on duration for each condition, as given by the coefficients a1 and a2, dimensionless measures of "Difficulty" and Stress effects on Duration, isolated from each other. The results are summarized in Tables 6.1, 6.2, 7.1, and 7.2.

**Table 6.1** The coefficient a1, a dimensionless isolated measure of "difficulty" effect on Duration, for each measure of "difficulty."

| "difficulty" | Pleasant | UnPleasant | UnHedonic |
|---|---|---|---|
| Difficulty | 0.314 | -0.179 | 0.877 |
| (-Grade) | 0.221 | 0.037 | 0.639 |
| (-Expected_Gr) | 0.159 | 0.045 | 0.569 |

**Table 6.2** Statistical significance of the coefficient a1 (p, a probability of accepting a1=0)

| "difficulty" | Pleasant | UnPleasant | UnHedonic |
|---|---|---|---|
| Difficulty | 0.04 | 0.28 | 0 |
| (-Grade) | 0.23 | 0.94 | 0 |
| (-Expected_Gr) | 0.59 | 0.91 | 0 |

**Table 7.1** The coefficient a2, a dimensionless isolated measure of Stress effect on Duration for each measure of "difficulty."

| "difficulty" | Pleasant | UnPleasant | UnHedonic |
|---|---|---|---|
| Difficulty | 0.176 | 0.221 | -0.897 |
| (-Grade) | 0.254 | 0.094 | -0.444 |
| (-Expected_Gr) | 0.235 | 0.082 | -0.545 |

**Table 7.2** Statistical significance of the coefficient a2 (p, a probability of accepting a2=0)

| "difficulty" | Pleasant | UnPleasant | UnHedonic |
|---|---|---|---|
| Difficulty | 0.28 | 0.37 | 0 |
| (-Grade) | 0.19 | 0.73 | 0 |
| (-Expected_Gr) | 0.22 | 0.76 | 0 |

**Discussion**

The first fundamental result of the current report demonstrates that music affects performance.

Table 1 shows that the test performance as measured by Grades confirmed the hypothesis: Grades for Pleasant music condition are higher than for UnPleasant or UnHedonic conditions. These differences are statistically significant.

Similar differences in the past were called the 'Mozart effect' and eventually dismissed as short-term effect, non-specific to music. However, confirmation of our second hypothesis discussed below demonstrates that the effect of music on performance can be expected short-termed, yet might be related to fundamental psychological mechanism: overcoming the morbid consequences of CD.

The second fundamental result of the current report deals with the cognitive function, origin, and evolutionary causes of music: music helps overcoming morbid consequences of CD. Thinking, accumulating knowledge, and making choices involves CD, which causes stress. Thinking is stressful. This stress reduces time humans allocate to thinking. This conclusion is supported by the third column in Table 7.1: the coefficient a2 is negative, Stress reduces duration of tests. This effect is highly statistical significant, which is seen from the $3^{rd}$ column in Table 7.2. Whereas naively one could expect that more stressful tests should require more time, these results demonstrate that when the effect of difficulty is separated, the effect of stress is opposite from this naïve expectation. Stress reduces duration because stress is unpleasant and tolerating stress is difficult. If humans in their evolutionary development would not be able to overcome this morbid consequence of CD, human culture would not evolve to more knowledge and to ability for thinking.

These results confirm the previously discussed hypothesis (Perlovsky 2010, 2012): music evolved for helping to overcome this predicament. Pleasant music helped keeping in mind contradictory cognitions in stressful thinking. This is seen from the 1[st] column in Table 7.1: coefficient a2 for Pleasant music condition is positive. In other words with pleasant music students were able to tolerate stress and devote more time to stressful thinking. The 1[st] column in Table 7.2 shows that the values of the coefficient a2 for Pleasant condition are of low statistical significance (comparative to value a2=0), but the effects of Pleasant music were highly statistically significant when compared to UnHedonic condition. In other word the effects of Pleasant music were highly statistically significant in terms of enabling stressful thinking.

The effect of UnPleasant music condition deserves future studies. The effect of stress on duration in UnPleasant condition is not statistically significantly different from a2=0 (no stress effect), and not significantly different from Pleasant condition. Still it is different from UnHedonic condition and this difference is highly statistically significant.

Evaluating results in Table 6.1, the effect of difficulty on duration, we would note that the mean value of the coefficient a1 is positive, 0.298, as well as its median value 0.221. As expected it is a positive value: difficulty increases duration (while a1 has low statistical significance, except one case, of subjective Difficulty, p=0.04). In UnHedonic condition coefficients a1 are highly statistically significant (p=0): more difficult tests take longer to solve. Linear correlation of difficulty and duration over conditions (0.163) is of low statistical significance because the relation is not linear, difficulty and stress have opposite effects on duration.

It must be underlined that the observed performance improvement, and therefore the usefulness of music, was present only with agreeable music and that unpleasant music tended to produce results that were often no different from controls. Such a result thus give a new evidence of the role of pleasurable experience in decision making (Cabanac & Bonniot-Cabanac, 2011; Ovsich & Cabanac, 2012; Perlovsky, Bonniot-Cabanac, & Cabanac, 2010; Bonniot-Cabanac, Cabanac, & Perlovsky, 2010; Ramírez et al, 2009; Bonniot-Cabanac & Cabanac, 2009; Cabanac & Bonniot-Cabanac, 2007; Ramírez, Bonniot-Cabanac, & Cabanac, 2005; Cabanac et al, 2002).

Studying three groups of students according to pleasant, unpleasant, and indifferent reaction to music is of course a first step. From theoretical arguments (Perlovsky, 2006, 2008, 2010a,b, 2011, 2012a,b) one could expect that there are very many musical emotions that help overcoming different CD emotions. Does any pair of contradictory cognitions cause a different CD emotion? Does any musical phrase contain a different emotion? Whereas experiments reported here should be reproduced as a matter of confirming our results, the directions of research should be expanded. Future research should develop experimental means of measuring musical emotions as well as CD emotions, establish relations of musical and CD emotions, and their relations to basic emotions (Perlovsky, Bonniot-Cabanac, & Cabanac, 2010; Fontanari et al 2012).

Music could be fundamental to the human ability to accumulate knowledge, to overcome irrational decision-making caused by CD, and to sustain the human cultural evolution in the face of ever increasing pressure from CD to devalue knowledge. This is related to the fact that useful knowledge contradicts instinctual drives[1], otherwise the instinctual drive would be sufficient and no knowledge would be needed. The same argument applies to any two elements of knowledge. Thus knowledge implies CD. So let us repeat, accumulation of knowledge and ability to think requires overcoming CD tendency to devalue knowledge.

It is interesting to note that Ancient Greeks knew about CD and the human tendency to devalue contradictory cognitions. In the Aesop's fable The Fox and the Grapes a fox sees high-hanging grapes. A desire to eat grapes and inability to reach them are in conflict. The fox overcomes this cognitive dissonance by deciding that the grapes are sour and not worth eating. Since the 1950s cognitive dissonances became a wide and well studied area of psychology. Let us repeat that tolerating cognitive dissonances is difficult, and people often make irrational decisions to avoid contradictions (Festinger, 1957; Tversky & Kahneman, 1974). In 2002 this research was awarded Nobel Prize in economics emphasizing the importance of this field of research. Nevertheless the psychological status of CD emotions have not been addressed: are these emotions similar to basic emotions, such as fear or rage, or are they fundamentally different (Perlovsky, Bonniot-Cabanac, & Cabanac, 2010)? Are there few CD emotions, similar to basic emotions (Shaver et al 1987; Russell, 1989; Cabanac, 2002; Petrov, Fontanari, & Perlovsky, 2012), or is there a virtual infinity of emotions (Cabanac, 2002), a very-high dimensional emotional space corresponding to every pair of cognitions (Perlovsky 2010; Fontanari et al, 2012)?

The reported research has demonstrated that unhedonic students, reporting no emotions from music, have scored lower grades. This is interesting in itself, but it raises a deeper psychological and anthropological question in view of our main hypothesis that music is crucial for the entire human cultural evolution. This question is: how large should be cognitive differences between musical and amusical people? (A significant percentage of people, about 1 per 100, are amusical, reporting no emotions experienced during listening to music, Groeger, 2012). No doubt, this is an interesting question for future experimental studies. Here we would suggest a theoretical hypothesis that the main contribution of music to culture could be in creating musical emotions, overcoming CD devaluations of knowledge, and sustaining human cultural evolution. As knowledge, cognition, and culture have evolved and are categorized in language, amusical people can participate in this cultural process. If insensitivity to music affects cognitive differences between musical and amusical people, these differences should be searched in specifically creative aspects of cognition, such as differences between decisions made using the knowledge instinct and using language-based heuristics (Perlovsky & Levine, 2012).

The "Mozart effects" was reported to be non-specific to music (Steele et al 1999; Thompson et al 2001; Schellenberg, 2006). It would be interesting to establish which nonmusical activities help overcoming CD. This paper explored two important and fascinating areas of human mind: music, its cognitive function, its origin, and cognitive dissonances. Each area deserves studying its deep multifaceted cognitive mechanisms and functions.

Our results confirmed the fundamental role of pleasure in decision making. These results are extremely important to all teachers as pleasant music improves academic performance, a topic that haunts any teacher, since suggested by the "Mozart effect" and then denounced as short-term effect nonspecific to Mozart or music (Tomatis, 1991; Wikipedia, 2012; Steele et al 1999;

---

[1] Here instinctual drives are defined according to (Grossberg and Levine, 1987) as inborn sensor-like neural mechanisms measuring vital needs of an organism.

Thompson et al 2001; Schellenberg, 2006). Yet we must also be aware that the unpleasant music might have had bad influence on academic performance, but the similar results obtained from UnHedonic and UnPleasant participants contradicts such a conclusion.

Let us repeat that our hypotheses stimulating this experiment have been that CD are implicit in any test and have a major influence on performance, duration, and stress, and Pleasure has a fundamental role in decision making and overcoming negative effects of CD. Majority of people, including students taking the tests dislike contradictions in their knowledge, experience it as Stress, and do not want to keep it in the mind for long: more Stress less Duration[2]. This is reversed during the Pleasant music condition. *The fundamentally important result is that pleasant music helped tolerating stress for longer and resulted in better Grades. Pleasant music helps overcoming CD (stress) and helps keeping in mind contradictory cognitions.*

## Conclusion

The current paper contributes to understanding two unsolved problems in psychology. The first is the origin and evolution of music. Consciousness is much evolutionary older than human music (Cabanac et al, 2009). Why has music such a power over us and how could it emerge in evolution (if it did) (Perlovsky, 2010, 2012; Ball, 2008; Editorial, 2008)? Together with theoretical considerations (Perlovsky, 2010, 2012) and experimental evidence (Masataka & Perlovsky, 2012) the current paper makes a significant contribution toward solving the problem of origin and cognitive function of music. The second unsolved problem is overcoming cognitive dissonances in cultural evolution: the current understanding of CD suggests that at the time of human emerging from animal kingdom, language and knowledge could not have evolved, because CD would lead to devaluing knowledge.

If the 'Mozart effect' is due to CD and due to the pleasure of music hearing helping to overcome CD-related stress and devaluation of knowledge, this suggests a natural explanation for short-term value of the 'Mozart effect.' Our paper suggests that long-term exposure to music and sensitivity to musical emotions are likely to be important for cognitive abilities, but this should be a separate field of study from the short-term 'Mozart effect.'

The current paper adds evidence to the emerging theory that music evolved jointly with language for the purpose of overcoming the morbid consequences of CD (Perlovsky, 2010, 2012; Masataka & Perlovsky, 2012). Educators and teachers invest much effort to minimize the emotion of examinations. The present results might lead to playing pleasant music in examination rooms.

## Acknowledgment

The authors wish to thank the students of De Rochebelle School (C.S.D.D., Quebec) who served as participants, and Dr. Roman Ilin for his kind statistical support. This work was supported by the Natural Sciences Research Council (NSRC) of Canada.

## References

---

[2] As known empirically by any teacher